\documentclass[10pt,conference]{IEEEtran}
\usepackage{amsthm, amsmath, amssymb, multirow, cite, comment, bm, array, mathrsfs} %graphics
\usepackage{graphicx , color} %wrapfig, subfig, 
\usepackage{caption}
\usepackage{float}
\usepackage{tikz} 
\usepackage[ruled,vlined]{algorithm2e}
\usepackage{tabularx}
\graphicspath{{Images/}} 
\usepackage{pict2e}
\usepackage{booktabs}  % professionally typeset tables
\usepackage{textcomp}
\usepackage{helvet}
\usepackage{graphicx,layouts}
\usepackage{caption}
\usepackage{float}
\usepackage{fullpage }
\usepackage{amsmath,amssymb,amsthm,enumitem}
\usepackage{amsfonts}
\usepackage{verbatim}
\usepackage{algpseudocode}
 \usepackage{wrapfig}
 \usepackage{subcaption}
\usepackage{authblk}
\usepackage{adjustbox}
\usepackage[ruled,vlined]{algorithm2e}

\begin{document}
\title{ \Large \textbf{ A Secret Key Generation Scheme for Internet of Things using Ternary-States ReRAM-based Physical Unclonable Functions} }

\author{Ashwija Reddy Korenda, 
Fatemeh Afghah, Bertrand Cambou}
\affil{School of Informatics, Computing and Cyber Systems, Northern Arizona University, Flagstaff, AZ 86011}
\renewcommand\Affilfont{\small}
\affil{\textit{\{ashwijakorenda,fatemeh.afghah,bertrand.cambou\}@nau.edu}}
 \maketitle
  \thispagestyle{empty}
 
 \begin{abstract}
Some of the main challenges towards utilizing conventional cryptographic techniques in Internet of Things (IoT) include the need for generating secret keys for such a large-scale network, distributing the generated keys to all the devices, key storage as well as the vulnerability to security attacks when an adversary gets physical access to the devices. In this paper, a novel secret key generation method is proposed for IoTs that utilize the intrinsic randomness embedded in the devices' memories introduced in the manufacturing process. %Resistive Random-Access Memory (ReRAM) is a promising storage technology for IoT devices, noting its considerable low energy consumption and long lifetime. 
%Therefore, we deployed these on-board ReRAM of the IoT devices and developed 
A fuzzy extractor structure using serially concatenated BCH-Polar codes is proposed to generate reproducible keys from a ReRAM-based \emph{ternary-state} Physical Unclonable Functions (PUFs) for device authentication and secret key generation. %In the proposed key generation method using ternary ReRAM-PUFs, the bits that are prone to easily flip from $0$ to $1$ (or vice-versa) are blanked, as they are not reliable enough to produce stable 0’s and 1’s. 
%This method provides a practical solution for the most common concern in key generation using memory-based PUFs which is the reproducibility of keys in different physical conditions, since the proposed ternary PUFs are capable of generating stronger secret keys by producing reliable outputs that can be reconstructed. 
The main concern in deploying PUF-based key generation methods is the leakage of information about the secret keys from the publicly available helper data. The fuzzy extractor proposed in this paper ensures much less mutual information between the generated keys and the helper data. The experimental results show that our proposed scheme is capable of generating notably stronger keys compared to existing techniques, while utilizing a significantly lower number of helper data bits. The failure probability when a low complex Successive Cancellation decoder is implemented in the proposed fuzzy extractor structure is $10^{-8}$ which was further increased to $10^{-10}$ when a complex iterative belief propagation decoder was used.\footnote{This project is partially supported by Arizona Board of Regents under Grant \# 1003330.}

\end{abstract}

Index Terms-  Strong cryptographic key, Fuzzy Extractor, ReRAM, PUFs, Polar Codes.

\section{Introduction}
The emergence of Internet of Things (IoT) has led to a technology breakthrough, but it comes with a set of new challenges including security, radio spectrum scarcity, complexity of network, and data management. IoT nodes are prone to selfish and malicious attacks. The selfish attacks can be reduced by using a reputation based model in a property rights spectrum sharing model using coalition formation techniques\cite{korenda2017}. Once spectrum is allocated to the IoT nodes, we have to ensure that the the information is sent to, or received from an authenticated user. Authentication is currently based on key-based cryptographic mechanisms that heavily rely on uniformly distributed, robust, reliable, and reproducible secret keys which are stored in the non-volatile memory (NVM) of the device \cite{dodis2004fuzzy}. However, these NVMs are highly vulnerable to physical attacks due to their robust electrical nature and maintaining the secrecy and integrity of these keys is a strenuous task, particularly in large-scale heterogeneous IoTs. Protection against physical attacks often involves the use of active tamper detection and prevention circuitry that requires a constant power supply \cite{shamsi2016security}. Therefore, these approaches are expensive from both design area and power consumption perspectives. Hardware-based security mechanisms using Physical Unclonable Functions (PUFs), utilize the intrinsic variations introduced during the fabrication process which include the doping level of semi-conducting layers, threshold voltages, and critical dimensions, etc., to extract a ``fingerprint" of the device. This can be used for identification and authentication of the device and will allow us to eliminate the need to store the secret keys in the NVM of the devices \cite{fremanteau1999unforgeable, pappu2002physical,afghah2017reram}. 
\newline
PUF-based security mechanisms rely on building unique challenge and response pairs (CRPs) for each device that can be reproduced in different conditions. 
Strong PUFs are expected to offer a high enough inter Hamming distance (distance between responses of two different PUFs under same conditions) which makes each device unique from the other ones, while having a low intra Hamming distance (distance between two responses of a single PUF under same conditions) to reduce the amount of False Acceptance Rate (FAR) \cite{katzenbeisser2012pufs}.
However, the responses of PUFs are subject to variations due to temperature changes, aging, drifts, electro-magnetic interactions, and other sources of noise. To deal with such variations in the PUFs' responses in different physical conditions, several fuzzy extractors which use strong error correction coding (ECC) techniques have been proposed \cite{kang2014cryptographie,kang2013implementation,chen2017high,kang2014performance,delvaux2016efficient,taniguchi2013stable}. 
These conventional ECC schemes can provide an acceptable error correction capability where the PUFs are utilized for authentication purposes, since true authentication can be granted if the rate of matching responses is below a certain threshold \cite{cambouafghah2015}. 
However, these current ECC schemes deem insufficient when the PUFs are utilized for generating cryptographic keys, where even a single bit mismatch between the private keys generated from PUF responses in different conditions can invalidate the performance of the symmetrical encryption schemes such as Data Encryption Standard (DES) and Advanced Encryption Standard (AES). 
\newline
The contribution of this paper is to propose a novel key generation method using ternary state ReRAM-based PUFs to generate reproducible and reliable keys using less publicly available \textit{Helper Data}. Here, we propose a syndrome-based fuzzy extractor using serially concatenated BCH-Polar codes that can offer several unique advantages compared to current technologies:
\begin{itemize}
%\item{ Utilizing the ternary PUFs results in generation of much more reliable keys compared to conventional binary ones by only using the stable 0's and 1's.} 
\item{ Uses polar codes which can significantly reduce the mutual information between the generated key and the publicly available helper data compared to other schemes, and hence reduce the capability of attackers to reconstruct the key by using the helper data. }
\item{ Using serially concatenated BCH -Polar codes can exactly regenerate the key using helper data which is highly demanded in PUF-based key generation applications.}

%offer increase the entropy of the generated key from $2^n$ to $3^n$, where n is the number of bits in a key}
\end{itemize}
The rest of this paper is organized as follows: In Section \ref{sec:PUF}, a brief review of the state of the art PUF technologies, fuzzy extractor schemes and importance of using Polar codes in Fuzzy extractors is provided. In Section \ref{proposed}, the proposed fuzzy extractor scheme using serially concatenated BCH-Polar Codes is provided. Section \ref{results} discusses the experimental results and the performance of the proposed method.
\begin{figure*}	\vspace{-1.5 cm}
\includegraphics[width=\textwidth]{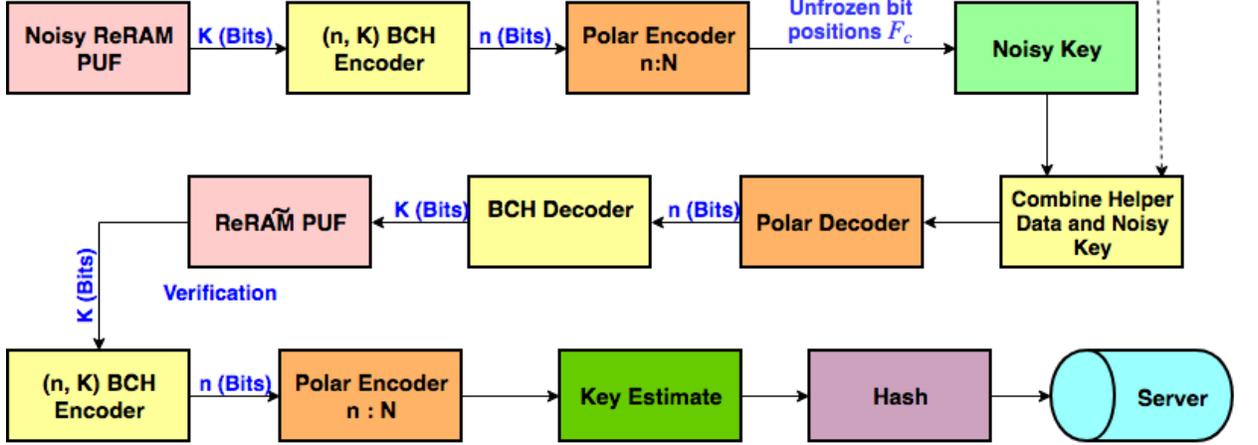} 
    \caption{Block diagram of the proposed key generation scheme for Key registration and Regeneration.}
    \label{fig: BlockDiagram}
    \vspace{-0.6 cm}
\end{figure*}

\section{Related Work}\label{sec:PUF}
\subsection{Memory-based PUFs}
Memory-based PUFs use the memory present in a chip/device to extract a fingerprint and therefore, they can be implemented in simple IoT devices with little or no extra parts added, to allow authentication in a network. 
%The resistance (response) is read from the memory cell after a biased current (challenge) is applied to it and the cell is programmed to either "0" or "1" based on the threshold. %\textcolor{red}{Cells in the memory device are in an unstable state (power up) and should be allowed to settle to a preferred stable state before extracting keys}. \textcolor{blue}{not very clear, better to say the challenge is the biased voltage and the response is read as such }
Several memory technologies have been deployed to build PUFs. SRAM-based PUF uses CMOS architecture and can be easily integrated in a system, but they give off energy while switching states that can be detected using a signal analyzer. Therefore, this type of PUF gives the hackers enough information to clone the device and extract the key\cite{helfmeier2014physical, ReRAMternaryPUF}.
\newline 
New IoT technologies often utilize more advanced memory devices such as Resistive Random Access Memory (ReRAM), memristive devices, and MRAM. Memristor and ReRAM rely on resistive technology to store information, where the current or the resistance of the cell when a biased voltage is applied can be used as a 	challenge \cite{pavan1997flash,cambou2015physically}. These devices have high randomness which not only occur in separate dies but also in the same die. Therefore, they can be a good candidate for a PUF. % if a good error correction technique is in place to reduce the probability of false negative authentication (FNA) rate when these devices are used in different physical conditions. %\textcolor{red}{Only specific parameters in the device can be measured with consistency such as low and high state resistance which can be used to represent "0"s and "1"s}.\textcolor{blue}{do you need this?}
 ReRAM based PUFs are faster when compared to MRAM or Flash and operate at or below noise level making them more resistant to side channel attacks without direct access to the chip. Operating power of ReRAM is around 10pJ/bit which is much lower compared to 1mJ/bit for Flash and 100pJ/bit for MRAM. 
%Moreover, the main property of ReRAM-based PUFs that make them an attractive choice for authentication and key generation in IoTs compared to flash memory is that they are faster and operate at or below noise level, therefore they are more resistant to side channel attacks without direct access to the chip. Operating power of ReRAM is around 10pJ/bit which is much lower compared to 1mJ/bit for Flash and 100pJ/bit for MRAM. 
\newline
While the accessibility, easy process of challenge and response generation, and operation at or below noise level in ReRAMs are of great interest to generate ideal PUFs, these memory devices are highly sensitive to the physical conditions (e.g. temperature, operating current, and electromagnetic interference) of the test environment. Besides such environmental circumstances, other factors including aging of the device and random noise sources can result in flipping of the programmed bits one way or other during the response generation. This in turn increases the probability of false negative authentication (FNA) when a single device is being authenticated in different conditions. Such errors can be partially corrected using ECC and fuzzy extractor schemes to allow the use of memory-based PUFs for authentication purposes. However, in key generation applications where the extracted key is used to encrypt the messages of a user, even a single bit error does not allow the decryption of the message. Thus limiting the applications of memory-based PUFs as cryptographic primitives \cite{yu2011lightweight}.
%This could result in 5-20\% CRP matching error rate if the ratio of marginal cells is high.
A new  concept of \emph{ternary PUFs} was proposed in \cite{ReRAMternaryPUF} to enhance the reliability of PUFs' responses by identifying three possible states of ``0", ``1" and ``X" for each cell rather than a conventional binary assignment. In the proposed ternary PUFs, the cells that are too close to the threshold and are not reliable are blanked by an ``X". This will allow reduction in CRP error compared to a binary scenario but, the CRP error is still not equal to 0. Here we utilize these ternary ReRAM-based PUFs to propose a practical solution for key generation described in Section \ref{proposed} to allow exact reproduction of the key.
%PUFs based on ReRAM'S have a linear relationship with current. Under different current conditions, even though the resistance measured in each cell is different from initially current conditions, these cells tend to lie in the same region of "0", "1" and "X" as before. 

\subsection{Fuzzy Extractors}
\textit{Fuzzy extractors} were proposed in \cite{dodis2004fuzzy} to convert noisy non-uniform PUF data into reproducible uniform strings. Fuzzy extractors can extract a uniformly random string $S$ and a non-secret helper data string $P$ from its original input $W$. This $P$ will allow exact reproduction of the string $S$ from noisy input data $W'$ which remains close to $W$.  %Fuzzy extractors are constructed using \textit{Secure Sketches}, which is a pair of randomized procedures ``sketch'' and ``recover'' which allow precise reconstruction of the noisy input by using helper data $W$ \cite{dodis2004fuzzy}.
A review of several constructions for fuzzy extractors are provided in \cite{dodis2004fuzzy}. Fuzzy extractor structures are based on utilizing different ECC techniques, among which BCH codes have been widely utilized \cite{dodis2004fuzzy,kang2014cryptographie,kang2014performance,kang2013implementation, delvaux2016efficient, taniguchi2013stable}. 
Two constructions for fuzzy extractors namely: code offset and syndrome-based constructions were proposed in \cite{dodis2004fuzzy}. In code-offset construction; helper data is generated by XORing the PUF input and the respective ECC codeword of the PUF input, while in the syndrome-based construction; helper data is generated by multiplying the PUF data with a parity matrix. Our proposed fuzzy extractor follows the syndrome-based construction, but can be extended to a code-offset construction by masking the PUF data using the ECC codewords.

%Many researchers had proposed different fuzzy extractor architectures in the literature whose results were compared in the Table \ref{fuzzyTable}.

\subsection{Polar Codes}
Polar codes are linear block ECCs whose construction is based on multiple recursive concatenation of short kernel codes which transforms the physical channel in to virtual outer channels whose capacity is either 0 or 1 \cite{arikan2009channel}. %Polar codes are based on the concept of channel polarization\cite{arikan2009channel}, where multiple recursive channel concatenation of short kernel codes can transform the capacity of $N$ independent copies of the channel to either $0$ or $1$. 
The channels whose capacity is 1 are called \textit{unfrozen channels} and they are used to sent information. These bits through which information is sent are called \textit{unfrozen bits}. The channels whose capacity is 0 are called \textit{unfrozen channels} and usually a 0 is sent through these channels. These bits through which no information is sent are called \textit{frozen bits}.
\newline
The position where the unfrozen and frozen bits appear during channel polarization is dependent on the \textit{Bhattacharya parameter} (measure of reliability) which is a function of the channels design parameters. 
%This property of Polar codes enables us to add another level of security to the key generated. 
Noting the considerable error correction capability of polar codes and more importantly, the low (almost zero) mutual information between the frozen and unfrozen bits in such codes, they can be utilized in combination with fuzzy extractors for key generation purposes. %used in combination with fuzzy extractor technique to generate strong cryptographic keys with less or zero mutual information between key and helper data by using unfrozen bits as the secret key and frozen bits as the helper data.  
In \cite{chen2017high}, a fuzzy extractor structure only based on Polar codes for binary SRAM PUF was proposed in which the key was generated using unfrozen bits of polar codes and the frozen bits were used as the helper data. This paper utilized a computationally complex successive cancellation decoder called Hash Aided SC List decoder (HA-SCL) to achieve a failure probability of $10^{-9}$ but utilized 896 bits to regenerate a 128 bit key. Thus utilizing data almost 7 times the size of the key for key regeneration. 

\section{Proposed Key Generation Scheme for Ternary ReRAM-based PUFs}\label{proposed}
Here, we propose a secret key generation scheme using ternary ReRAM-based PUFs, to generate unique keys from noisy PUFs' data by using less helper data compared to previously proposed schemes. A fuzzy extractor using a serial concatenation of BCH and polar codes is designed to generate the \textit{Secret Key} and \textit{Helper Data}. Using polar codes in the proposed scheme results in reducing the level of mutual information between the generated secret key and the publicly made available helper data, since there is almost zero mutual information between the frozen bits and unfrozen bits \cite{chen2017high}. Our proposed serially concatenated BCH-Polar model requires a lower number of helper data bits while offering a comparable failure probability compared to common fuzzy extractor schemes proposed in the literature as shown in the Table \ref{fuzzyTable}. This is possible as the BCH encoder also helps correct a few bits of the noisy PUF data, thus increasing the coding efficiency while having a low complexity. 
\newline
The stable 0's and 1's are obtained from the ReRAM PUF data, by first dividing the ReRAM PUF data into three states ``0", ``1", and ``X" and then discarding the memory cells which are prone to flip due to slight variations in environment (X) \cite{cambou2017ag}. This will essentially extract a stable binary PUF from the ternary PUF which can be used to generate strong, reliable and reproducible cryptographic keys \cite{afghah2017multi}.

% The resistance for each Re-RAM cell was measured when a certain current is applied. These resistance measurements were taken repeatedly for about a 100 times and the resistance data was recorded. Observing this data, the measurements were mapped to three states ``1'', ``0'' and ``X''. Discarding the measurements with ternary ``X'' state and considering only the stable ``0''s and ``1''s, strong reproducible PUF data can be produced.
 \begin{figure*}[h]
        \begin{subfigure}[b]{0.5\textwidth}
                \centering
                \includegraphics[width=.65\linewidth]{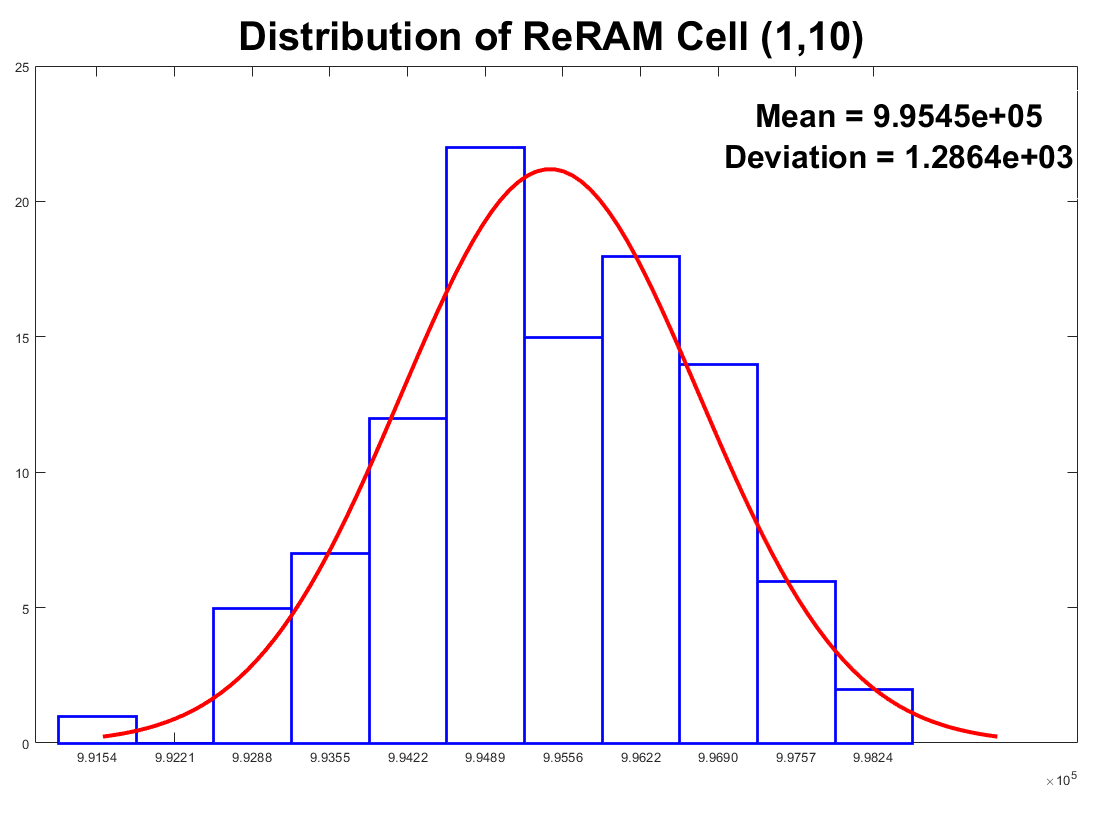}
        \end{subfigure}%
        \begin{subfigure}[b]{0.5\textwidth}
                \centering
                \includegraphics[width=.65\linewidth]{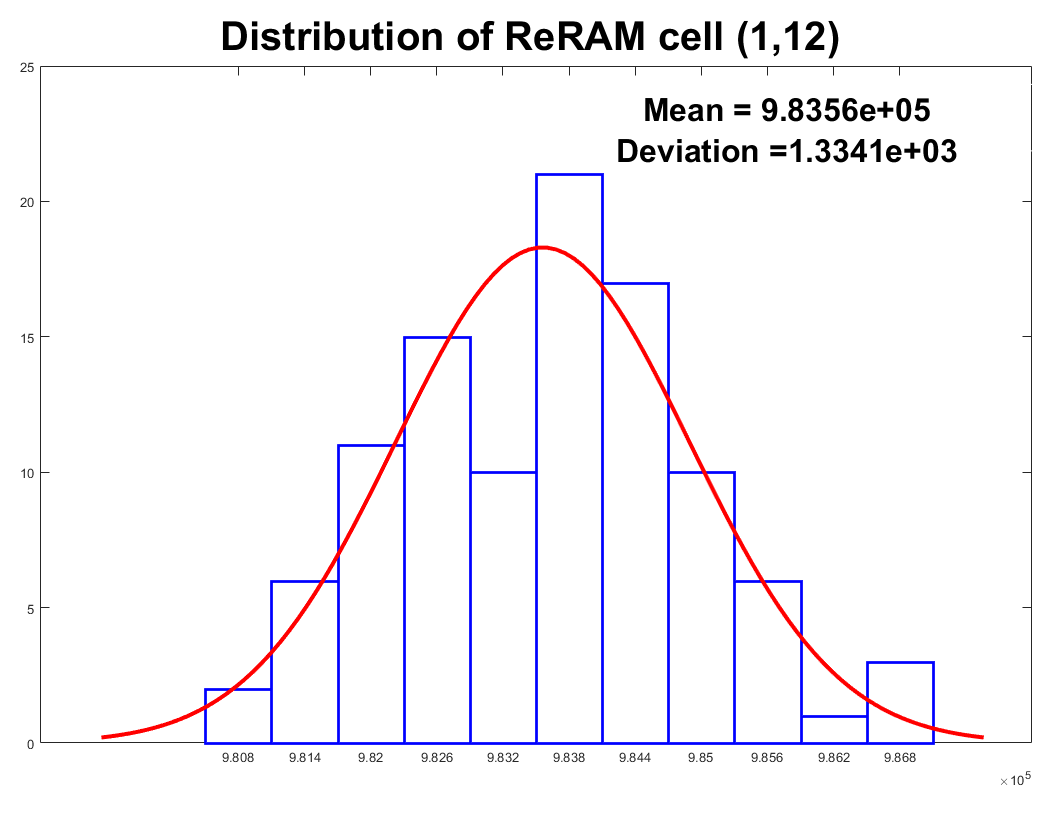}

        \end{subfigure}%
         \vskip\baselineskip
        \caption{Distribution of various Re-RAM Cells}\label{fig:DistributionReRAM}
        \vspace{-0.6 cm}
\end{figure*}

\vspace{-0.1 cm}
\subsection{Registration}
\begin{figure}
\includegraphics[width=\columnwidth]{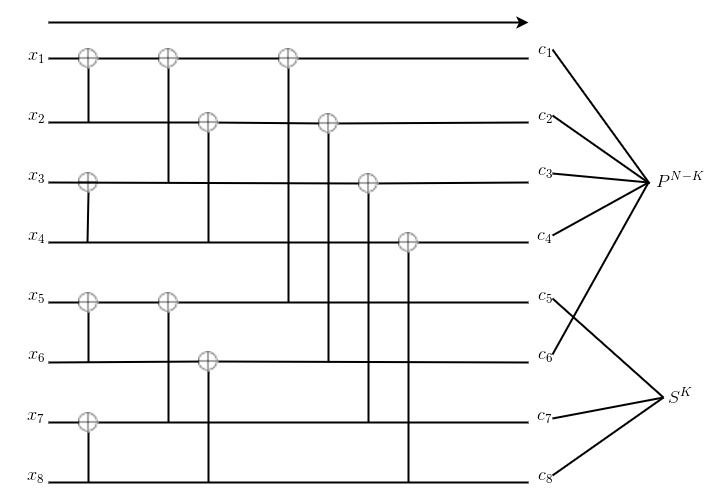} 
    \caption{Polar Encoder to select Key and Helper Data from BCH encoded ReRAM PUF.}
    \label{Fig: ExamplePolarEncoder}
    \vspace{-0.5 cm}
\end{figure}

In the \textit{Registration} phase, the ReRAM PUF data is encoded using a BCH encoder to increase the coding gain and allow reconstruction of the PUF responses with minimum error and less helper data. The BCH encoded output is then sent to a polar encoder whose output is unique for every input $U^N = X_{BCH}^NG^N$, where $U^N$ is the polar encoded output, $X_{BCH}^N$ is the BCH encoded PUF data, and $G^N$ is the generator. The bits that occur in the non-frozen positions at the output of the polar encoder are hashed and stored in the server as a \textit{secret key}, $S^K$ and the \textit{helper data}, $P^{N-K}$ is obtained from the frozen bit positions of the polar encoded data. Helper data and secret key can be represented by the equations below,
\begin{equation} 
P^{N-K}\cong X_{BCH}^NG^N_{\mathcal{F}} = C^N[\mathcal{F}]
\end{equation}
\begin{equation}
S^K \cong X_{BCH}^NG^N_{\mathcal{F}_\mathcal{C}}= C^N[\mathcal{F}_{\mathcal{C}}]
\end{equation}
where $\mathcal{F}$ represents the frozen bit indices and $\mathcal{F}_C$ represents the non-frozen bit indices. As the cells are assigned 0's and 1's based on their individual resistances, the cells are independent of each others behavior. An example of a Polar Encoder used in Key Generation scheme with \{1,2,3,4,6\} frozen bits representing five Helper Data bits made publicly available to regenerate the three bit key using Noisy PUF data is shown in Fig. \ref{Fig: ExamplePolarEncoder}. In this figure, $x_i$ represents the BCH encoded PUF bits and $c_i$ represents the polar encoded output of $x_i$.  
\begin{table*}
[ht]\caption{Comparison Of different fuzzy extractor schemes proposed in the literature.}
 \label{fuzzyTable}
\begin{center}
\begin{tabularx}{\textwidth}{|c|c|c|c|c|}
			\hline
			Fuzzy Extractor Construction   & Key Length    & Helper Data bits           & Failure Probability  & Flipping probability          \\ 
			\hline
			Reed Muller Generalized Multiple Concatenated coding \cite{maes2009low}	  &128    & 13952   & $10^{-6}$ & 15\%  \\ 
			\hline
			BCH Repetition Code	 \cite{maes2012pufky}   &128  & 2052   & $10^{-9}$ & 13\%      \\ \hline
			Generalized Concatenated (GC) Reed Muller \cite{puchinger2015error} 		& 2048     & 2048   & $5.37. 10^{-10}$ & 14\%     \\
			 \hline
			GC Reed Solomon \cite{puchinger2015error}                        &1024 & 1024 &  $3.47. 10^{-10}$ & 14\% \\
			 \hline
			Polar Codes SC \cite{chen2017high}      &128                 & 896             &  $ 10^{-6}$     &15\%       \\ 
			\hline
			HA SCL Polar Codes \cite{chen2017high}      &128                  & 896             &  $ 10^{-9}$     &15\%       \\ 
			\hline
			{\bf Proposed Key Generation Scheme using SC decoder}      & {\bf250 }                         & {\bf 262}          & $ {\bf 10^{-8
			}}$  & {\bf 15\%}        \\ \hline
            {\bf Proposed Key Generation Scheme using BP decoder}      & {\bf250 }                         & {\bf 262}          & $ {\bf 10^{-10
			}}$  & {\bf 15\%}        \\ \hline
        	  			\end{tabularx}
                        
\end{center}

\end{table*}

\subsection{Key Regeneration}
During \textit{Key Regeneration} phase, a polar decoder and BCH decoder are utilized to observe the output of the BCH-Polar encoded noisy PUF along with publicly available Helper Data. The Helper data is combined with the BCH-Polar encoded noisy PUF  by simply substituting the frozen bits of the noisy encoded data with the helper data bits. The Polar and BCH decoder utilizes this combined data to get a close estimate of the initial ReRAM PUF input. When a low complex SC decoder is utilized as a Polar decoder, its performance is not good enough for short block lengths, hence it can be further improved using SC list decoder, or a Belief propagation decoder. Fig. \ref{fig: BlockDiagram} summarizes the  construction of \textit{Key} and \textit{Helper data} using ReRAM PUF data.
\section{Results}\label{results}

\begin{figure}
\vspace{-0.5 cm}
\includegraphics[width= \columnwidth]{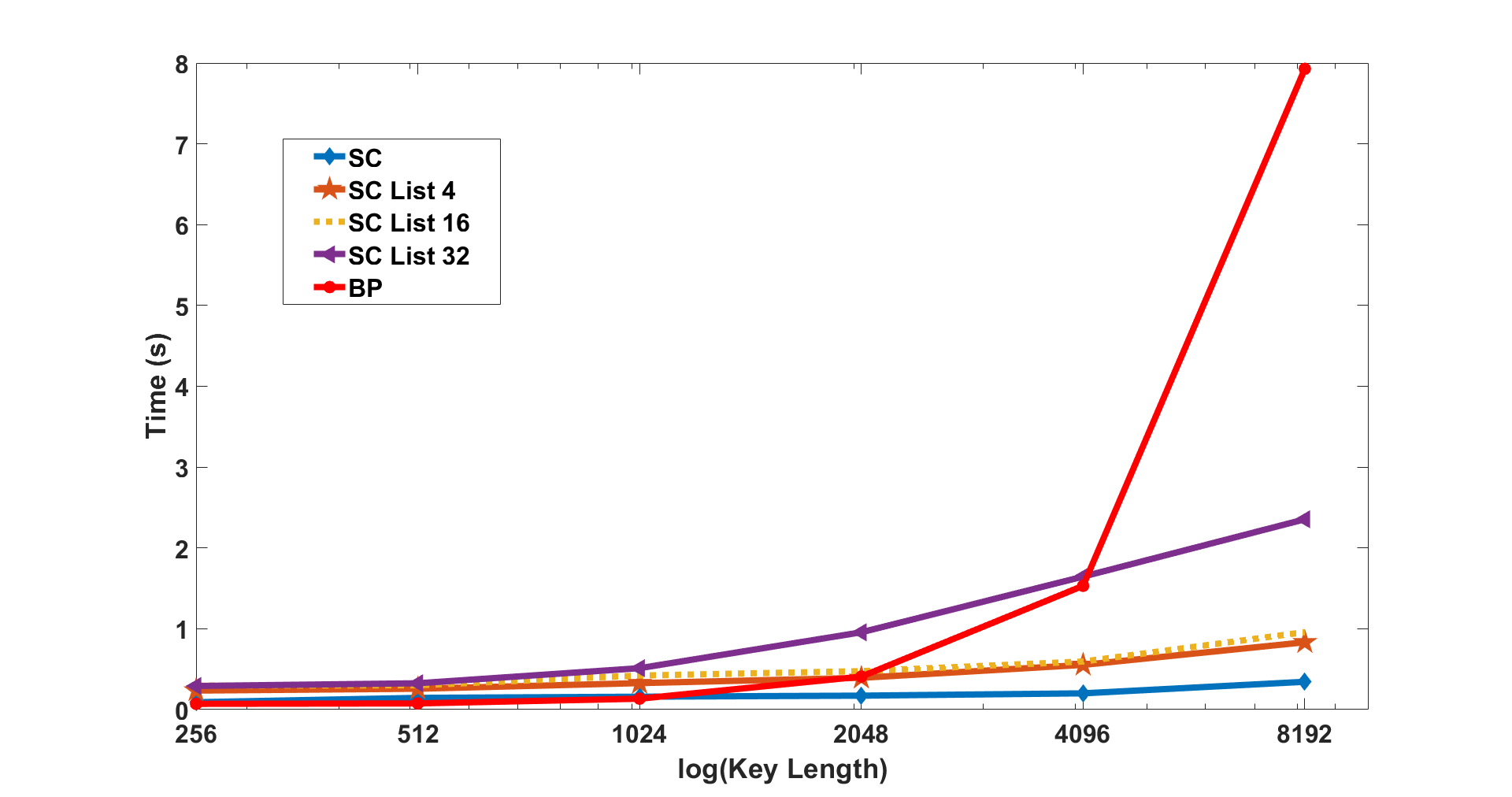}
\caption{Time complexity comparison of the Fuzzy extractor while using different Polar decoders.}
\label{Fig:timing}
\vspace{-0.5 cm}
\end{figure}

In this study, we use the samples from a ReRAM PUF fabricated at Northern Arizona University's CyberSecurity laboratory. This PUF comprises of a $HfO_2$ active solid electrolyte chalcogenide with a Hafnium based top electrode, and a tungsten based bottom electrode. The measured parameters from each cell include: i) stepping the voltage in the positive voltage direction at constant compliance current to measure the programing voltage, $V_{\text{set}}$; ii) after programming, measurement of the low resistive states $R_{\text{on}}$. The measurements from 254 fabricated ReRAM cells are available, where each measurement was repeated 102 times. After analyzing the data, the measurements of $R_{\text{on}}$ were used to generate the CRPs as it offers the desired performance metrics: i) the measurement of each cell is independent and is normally distributed, and ii) the readings of resistance for different cells are also independent and normally distributed. Fig. \ref{fig:DistributionReRAM} shows the independency and normal distribution of the $R_{\text{on}}$ measurements for two randomly selected cells. 
\newline
In order to determine the optimum threshold values to classify the cell responses to `0', `1', and `X', the distribution of all the 254 cells is plotted, and then range of the response distribution is divided into three equal parts using two thresholds, $Threshold_1$ and $Threshold_2$.
The cells whose resistance is below $Threshold_1$ (lies in the left most part of the histogram), are programmed with a `0'. The cells whose resistance values lie in between $Threshold_1$ and $Threshold_2$, are programmed with `X'. The cells with the resistance values higher than $Threshold_2$ are programmed with a `1'. 
\newline
\begin{figure}
 \includegraphics[width=\columnwidth]{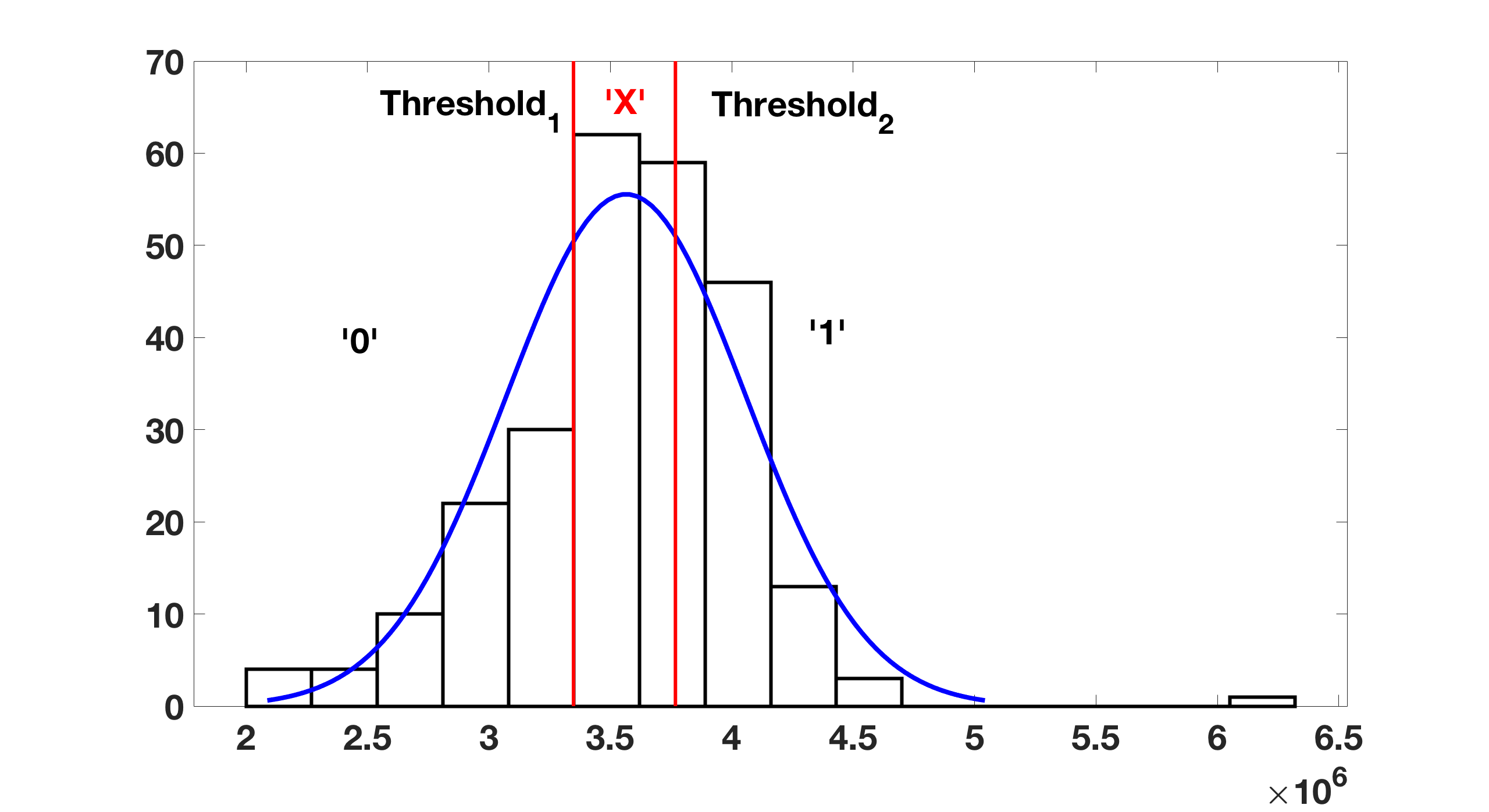}
 \caption{Distribution of 254 ReRAM cells and extraction of stable 0's and 1's using $Threshold_1$ and $Threshold_2$}
 \label{Fig:cellDistribution}
 \vspace{-0.5 cm}
 \end{figure}
The utilized ternary PUFs, exclude the unstable cells programmed with a `X' and do not used them in generating the secure keys'. Thereby, the binary output of the generated ternary PUFs are more stable and reliable compared to a conventional binary PUF. Figure \ref{Fig:cellDistribution} shows the distribution of the 254 cells and the thresholds used to divide the data into three states.
\newline
 \begin{figure}[hb]
 \includegraphics[width=\columnwidth]{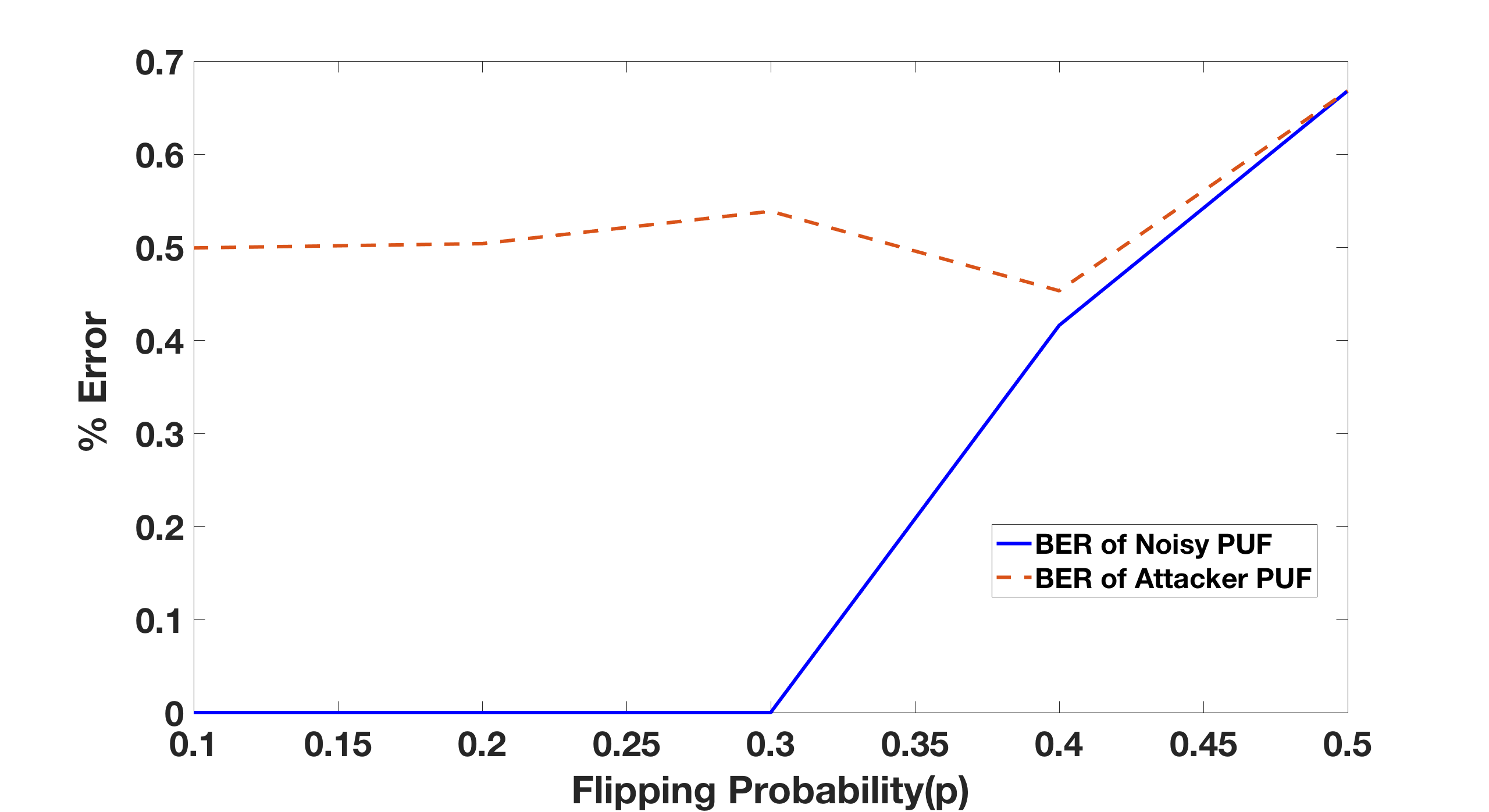}
 \caption{Percentage error between Key generated from Noisy or Attacker PUF and Key stored in the server}
 \label{Fig:BER}
 \end{figure}
\begin{figure}[hb]\vspace{-0.3 cm}
 \includegraphics[width=\columnwidth]{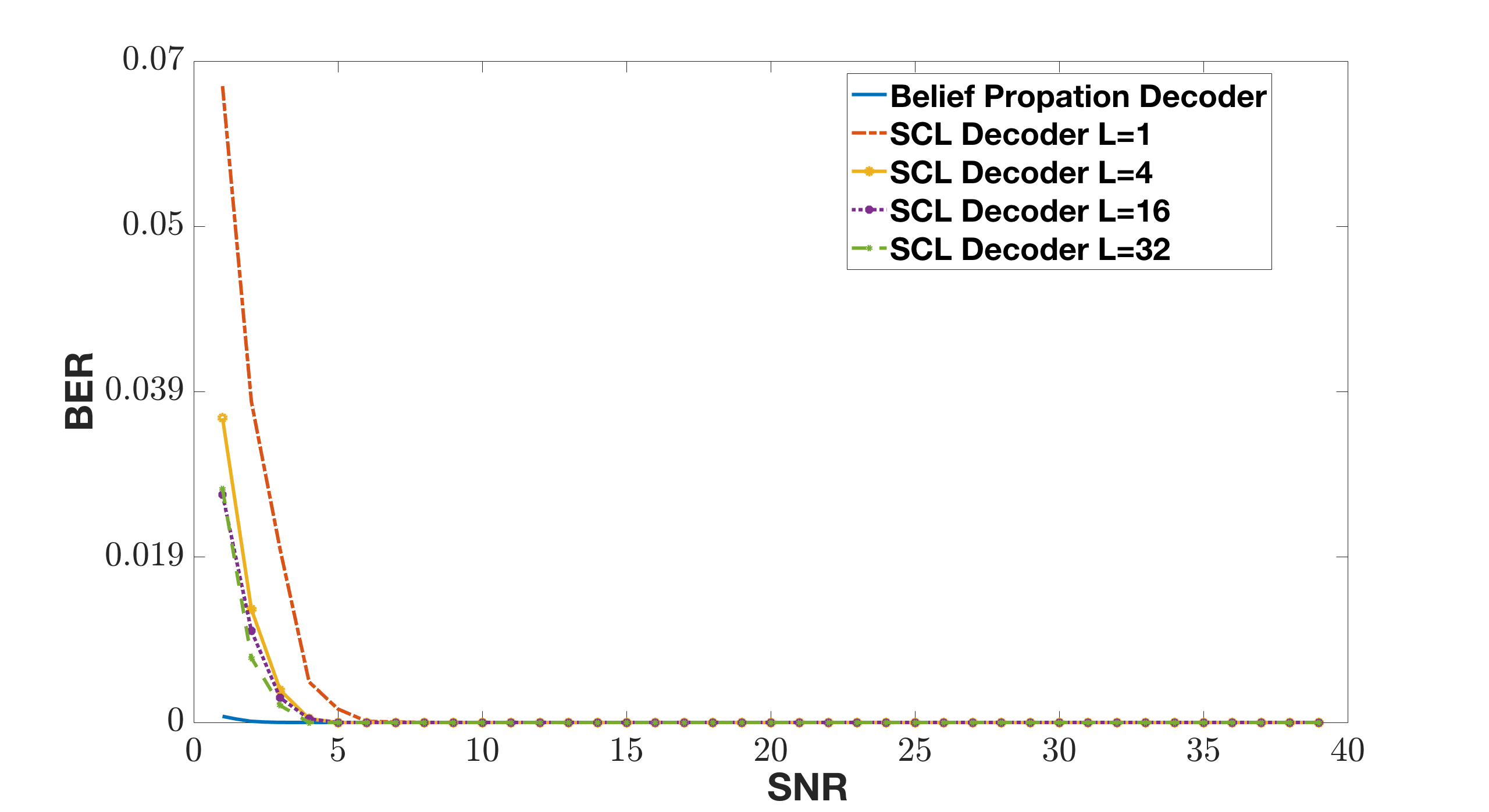}
 \caption{BER of the Regenerated Key while using different Polar Decoders}
 \label{Fig:BER}
 \end{figure}
In the Registration phase, the readings of $R_{\text{on}}$ for the available 254 cells were mapped to `0', `X' or `1'. A mask was used in order to extract 131 stable bits from the available 254 cells. These stable PUF bits are initially encrypted using a (255, 131) BCH encoder. Then, the 255 BCH encoded PUF data bits were encoded using a (512, 255) Polar Encoder. At the output of the Polar Encoder, the frozen bits (262 bits) were selected as \textit{Helper Data} while the unfrozen bits were hashed using a ``SHA-256" and stored in the server as the \textit{Secret Key}.
\newline
During the key regeneration phase, the repeated  data measurements from 254 cells were used to generate a \textit{noisy key} by selecting \textit{unfrozen bits} at the output of a Polar encoder. The \textit{Helper Data (262 bits)} that was extracted during the key generation phase, was combined with the \textit{noisy key} and given to a Polar-BCH serial decoder to generate an estimate of the initial 131 PUF data bits. This estimate is used to regenerate the \textit{Key} by giving it to a serially concatenated BCH-Polar encoder and selecting the \textit{unfrozen bits}, which is then hashed and compared against the \textit{Key} stored in the server.
\newline
The proposed method was also tested using simulated data in order to test the efficiency of the fuzzy extractor to regenerate keys. A noisy version of the initial data with a flipping probability of  15\% was used to regenerate the key. The fuzzy extractor could regenerate the key exactly with a failure probability of $10^{-8}$ while using a low complex Successive Cancellation decoder and $10^{-10}$ while using a iterative Belief Propagation decoder. 
\newline
The percentage of error in regenerating the \textit{Key} using \textit{Helper Data} for original PUF data under different simulated flipping probabilities was compared with an Attacker PUF with an inter-Hamming distance $\geq$ 40\% and the results were plotted in Figure \ref{Fig:BER}. The results indicated that, if the flipping probability is $\leq$ 30\%, then the proposed key generation scheme can exactly recover the \textit{Key} using \textit{Helper Data}. The attacker was not successful in recovering the original PUF data by using the publicly available \textit{Helper Data} and the percentage of error when the Attacker PUF is used to regenerate the key was about 50\%. 
\newline
The performance of different Polar Decoders was tested and plotted in Figure \ref{Fig:BER}, which shows that the Belief Propagation Decoder has a better performance compared to the SC and SC List decoders. This increase in performance comes with an increase in time complexity in regenerating the key which is compared in Figure \ref{Fig:timing}.
\newline
Table \ref{fuzzyTable} shows the comparison between the failure probability of key regeneration using publicly available helper data utilizing different fuzzy extractor schemes proposed in the literature.  As seen in this table, our proposed method results in a failure probability of $10^{-8}$ using a low complex Successive Cancellation decoder for key regeneration with a key length of $250$ bits while using $262$ bits of helper data, while the competitive methods require a large number of helper bits to be able to regenerate the keys with such a low probability of mismatch. This failure probability was further reduced to $10^{-10}$ when a complex Belief Propagation decoder was utilized.

\section{Conclusion}
The ReRAM based PUFs are the most practical choice for authentication and key generation in IoT, as they operate at or below the systems' noise level and therefore are less vulnerable to side channel attacks compared to the alternative memory technologies. However, the current ReRAM-based PUFs present a high false negative authentication rate since the behavior of these devices can vary in different physical conditions that results in a low probability of regenerating the same response in different attempts. 
In this paper, we propose a secret key generation scheme for ternary state PUFs that enables reliable reconstruction of the desired secret keys utilizing a serially concatenated BCH-Polar fuzzy extractor. The experimental results show that the proposed model can offer a significantly lower probability of mismatch between the original key and the regenerated ones, while a less number of \textit{Helper data} bits were used to extract the \textit{Key} when compared to previously proposed fuzzy extractor techniques.

{\footnotesize
\bibliographystyle{IEEEtran}
\bibliography{IEEEabrv}}
\end{document}